\documentclass[12pt]{revtex4}

\usepackage{hyperref}
\usepackage{amsmath,amsfonts,amssymb}
\hypersetup{dvips,dvipdfm,colorlinks=true,urlcolor=magenta,filecolor=magenta,linktoc=page,citecolor=red,linkcolor=blue,bookmarks=true}
\usepackage{graphicx,epstopdf}

\begin{document}

\title{Multiscalar field cosmological model and possible solutions using Noether symmetry approach}
\author{Santu Mondal$^1$\footnote {santumondal050591@gmail.com}}
\author{Roshni Bhaumik$^1$\footnote {roshnibhaumik1995@gmail.com}}
\author{Sourav Dutta$^2$\footnote {sduttaju@gmail.com}}
\author{Subenoy Chakraborty$^1$\footnote {schakraborty.math@gmail.com}}
\affiliation{$^1$Department of Mathematics, Jadavpur University, Kolkata-700032, West Bengal, India\\$^2$Department of Mathematics, Dr. Meghnad Saha College, Itahar, Uttar Dinajpur-733128, West Bengal, India.}

%%%%%%%%%%%%%%%%%%%%%%%%%%%%%%%%%%%%%%%%%%%%%%%%%%%%%%%%%%%%%%%%%%%%%%%%%%%%%%%%%%%%%%%%%%%%%%%%%%%%%%%%%%%%%%%%%%%%%%%%%%%%

\begin{abstract}
In this work, a cosmological model is considered having two scalar fields minimally coupled to gravity with a mixed kinetic term. The model is characterized by the coupling function and the potential function which are assumed to depend on one of the scalar fields. Instead of choosing these functions phenomenologically here, they are evaluated assuming the existence of Noether symmetry. By appropriate choice of a point transformation in the augmented space, one of the variables in the Lagrangian becomes cyclic and the evolution equations become much simpler to have solutions. Finally, the solutions are analyzed from cosmological view point. 
\end{abstract}

\maketitle

\section{Introduction}
Today, it is well accepted that Einstein gravity with normal matter is not sufficient to describe the present evolution of the Universe. To explain the present accelerated expansion as predicted by the series of observational evidences \cite{r1, o1, o2} for the last two decades, one needs some exotic matter having large negative pressure (known as Dark energy(DE)). Cosmologists started with cosmological constant as the DE candidate -- the simplest and observationally most favourable. However, due to two severe drawbacks of the cosmological constant, namely the fine-tuning problem and coincidence problem, the cosmologists are searching for an alternative choice which is commonly known as dynamical DE model. A quite well-known candidate for dynamical DE is the scalar field model (minimally or non-minimally coupled to gravity).  

Usually, in scalar field, cosmology quintessence model, phantom model and even multi-scalar field models are commonly used. In particular, multi-scalar field cosmology has a wide range of applicability such as an alternative models of inflation, i.e, hybrid inflation, $\alpha$--attractors and double inflation  \cite{c1, c2, c3, c4, c5, c6, c7, c8, c9}. A very common multi-scalar field model is the quintom model having two scalar fields of which one is the quintessence field while the other one is of phantom nature \cite{c10}. An existence of the quintom model having a mixed kinetic term has been introduced in Ref.\cite{c11}, considering the space of the kinetic energy to have flat geometry. A commonly used two-scalar field model having non-vanishing curvature of the two-dimensional (2D) space is known as chiral cosmological model \cite{c9, c12, c13}and has analogy with the nonlinear sigma cosmological model \cite{c14, c15, c16}.  \\
In this work, a two-scalar field model (for multi-scalar field models, see Refs.\cite{m1, m2, m3, m4, m5, m6, m7, m8, m9, m10, m11}) has been considered having one quintessence field and the other scalar field interacts with the quintessence field through the kinetic term. In order to have cosmological solutions of this complicated model, Noether symmetry approach has been introduced. The two unknown functions in the model, namely the potential function (which drives the dynamics of the quintessence) and the interaction term between the two fields are determined from the consistency conditions of the symmetry analysis, rather than choosing phenomenologically. This type of cosmological model (Chiral cosmological model) is actively studied by several authors. In particular, Noether symmetry approach has been proposed for the search of exact solutions in Chiral cosmological models by Paliathanasis and Tsamparlis \cite{palia}. Also there are other methods for the construction of exact solutions in Chiral cosmology \cite{a, 1}. This paper is organized as follows: Evolution equation of the model is presented in Sec.2, whereas Sec.3 deals with the Noether symmetry approach of this model. Sec.4 describes the analytical solutions and the paper ends with concluding remarks in Sec.5.

\section{Scalar Field Cosmology: Basic Equation}
The action of the scalar field $\phi(x^{\mu})$ and $\psi(x^{\mu})$ coupled to gravity has the form \cite{r2}
\begin{equation}
\mathcal{A}=\int d^4x \sqrt{-g}\mathcal{L}\bigg(R(x^{\mu}),\phi(x^{\mu}),\psi(x^{\mu}),\bigtriangledown_{_\nu}\phi(x^{\mu}),\bigtriangledown_{_\nu}\psi(x^{\mu})\bigg), \label{1}
\end{equation}
which is characterized by the Lagrangian density given by
\begin{equation}
\mathcal{L}=\frac{1}{2}R-\frac{1}{2}\bigtriangledown_{_\mu}\phi \bigtriangledown^{^\nu}\phi-\frac{F(\phi)}{2}\bigtriangledown_{_\mu}\psi \bigtriangledown^{^\mu}\psi-V(\phi),\label{2}
\end{equation}
where $8\pi G=1$ is the gravitational coupling, $F(\phi)$ and $V(\phi)$ are the coupling function and the potential function of the scalar field $\phi$, $R$ is the scalar curvature.\\

Now let us consider the flat FLRW space-time metric given by the line element
\begin{equation}
ds^2=g_{\mu \nu} dx^{\mu} dx^{\nu}=-dt^2+a^2(t)(dr^2+r^2 d\Omega_2^2),\label{3}
\end{equation}
where $a(t)$ represents the scale factor of the Universe, $d\Omega_2^2=d\theta^2+\sin^2 \theta d\phi^2$. So the above point like Lagrangian has the explicit form
\begin{equation}
L=-3a\dot{a}^2+\frac{1}{2}a^3\dot{\phi}^2+\frac{a^3}{2}F(\phi)\dot{\psi}^2-a^3 V(\phi).\label{8}
\end{equation}
 Then by varying the action with respect to the scale factor ``$a$'' and the scalar fields ``$\phi$'' and ``$\psi$'', the Einstein field equations take the form (Eq. (\ref{4}) is obtained by using the action with respect to the Lapse function \cite{b} chosen as unity in this context)
\begin{eqnarray}
\frac{1}{2}\bigg[F(\phi)\dot{\psi}^2+\dot{\phi}^2-6H^2 \bigg]+V(\phi)&=&0,\label{4}\\
\frac{1}{2}\bigg[-4\frac{\ddot{a}}{a}+4H-2H^2-F(\phi)\dot{\psi}^2-\dot{\phi}^2 \bigg]+V(\phi)&=&0,\label{5}\\
\frac{1}{2}\bigg[2\ddot{\phi}+6H\dot{\phi}-F'(\phi)\dot{\psi}^2\bigg]+V'(\phi)&=&0,\label{6}\\
\frac{d}{dt}\bigg(a^3F(\phi)\dot{\psi}\bigg)&=&0,\label{7}
\end{eqnarray}
where an overdot denotes the derivatives with respect to the cosmic time ``$t$'', ``$H$'' is the usual Hubble parameter defined by $H\equiv \frac{\dot{a}}{a}$ and the conserved quantity of the reduced system is given by (from (\ref{7}))
%\begin{equation}
$$Q:=a^3F(\phi)\dot{\psi}=m_1,$$
%\end{equation}
where $m_1$ is an integration constant.

\section{Existence of Noether symmetry}
Noether's first theorem states that every physical system is associated to some conserved quantities provided the Lagrangian of the system is invariant with respect to the Lie derivative \cite{n1, n2, n3, n4} along appropriate vector field ($\mathcal{L}_{\vec{X}}L=\vec{X}L$). Further these symmetry constraints help the evolution equations of the physical system to be solvable or to be simplified to a great extent. 

For a point like canonical Lagrangian 
\begin{equation}
L=L\bigg[q^{\alpha}(x^i)~\partial_j q^{\alpha}(x^i)\bigg],
\end{equation}\label{n1}
with the generalized coordinates $q^{\alpha}(x^i)$, the usual Euler-Lagrangian equations i.e.,
\begin{equation}
\partial_j\bigg(\frac{\partial L}{\partial~\partial_j q^{\alpha}}\bigg)=\frac{\partial L}{\partial q^{\alpha}},~~\alpha=1, 2.....,N
\end{equation}\label{n2}
can be contracted to some unknown function $\lambda^{\alpha}(q^{\beta})$ as
\begin{equation}
\lambda^{\alpha}\bigg[\partial_j\bigg(\frac{\partial L}{\partial~\partial_j q^{\alpha}}\bigg)-\frac{\partial L}{\partial q^{\alpha}}\bigg]=0,
\end{equation}\label{n3}
which can be written as \cite{n5}
\begin{equation}
\mathcal{L}_{\vec{X}}L={\vec{X}}L=\lambda^{\alpha}\frac{\partial L}{\partial q^{\alpha}}+\bigg(\partial_j\lambda^{\alpha}\bigg)\frac{\partial L}{\partial~\partial_j q^{\alpha}}=\partial_j\bigg(\lambda^{\alpha}\frac{\partial L}{\partial~\partial_j q^{\alpha}}\bigg).\label{n4}
\end{equation}
Here \begin{equation}
	\vec{X}=\lambda^{\alpha}\frac{\partial}{\partial q^{\alpha}}+\bigg(\partial_j~\lambda^{\alpha}\bigg)\frac{\partial }{\partial~\partial_j q^{\alpha}},\label{n5}
\end{equation}
is the vector field in the augmented space (the space consists of the dependable variable and their derivatives. Here in the present problem, it is a 4D space $(t, a, \phi, \psi)$). However if the above vector field $\vec{X}$ is the infinitesimal generator of Noether symmetry, then Noether's theorem tells us that $\mathcal{L}_{\vec{X}}L=0$. As a consequence from equation (\ref{n4}), there corresponds a conserved current associated with this symmetry, namely \cite{n6}
$$Q^{i}=\lambda^{\alpha}\frac{\partial L}{\partial~\partial_i q^{\alpha}},$$
with $\partial_i Q^i=0$.\\
Further, the Hamiltonian i.e., the energy function, is a constant of motion of the system provided there is no explicit time dependence in the Lagrangian. Moreover, if the conserved quantity due to the symmetry has some physical analogy \cite{n7} then the Noether symmetry approach can be used to identify the reliable model. In this context, the application of Noether symmetry is two-fold--to simplify the evolution equations and to solve the physical problem exactly.

In this section for solving the field equations (\ref{4}--\ref{6}), we use the above symmetry approach. Then according to this symmetry, a Lagrangian admits Noether symmetry if there exist a vector-valued function $G(t, a, \phi, \psi)$ which satisfies \cite{r3, r4}
\begin{equation}
X^{[1]}L+L D_{_t} \xi(t, a, \phi, \psi)=D_{_t} G(t, a, \phi, \psi),\label{11}
\end{equation}
under the vector field
$$X=\xi(t, a, \phi, \psi) \frac{\partial}{\partial t}+\alpha(t, a, \phi, \psi)\frac{\partial}{\partial a}+\beta(t, a, \phi, \psi)\frac{\partial}{\partial \phi}+\gamma(t, a, \phi, \psi)\frac{\partial}{\partial \psi},$$
where the total derivative operator $D_{_t}$ is given by $D_{_t}\equiv \frac{\partial}{\partial t}+\dot{a}\frac{\partial}{\partial a}+\dot{\phi}\frac{\partial}{\partial \phi}+\dot{\psi}\frac{\partial}{\partial \psi},$ and $X^{[1]}$ is the first prolongation vector given by
\begin{equation}
X^{[1]}=X+(D_{_t}\alpha-\dot{a} D_{_t}\xi)\frac{\partial}{\partial \dot{a}}+(D_{_t}\beta-\dot{\phi} D_{_t}\xi)\frac{\partial}{\partial \dot{\phi}}+(D_{_t}\gamma-\dot{\psi} D_{_t}\xi)\frac{\partial}{\partial \dot{\psi}},\label{12}
\end{equation}
and the conserved quantity associated with the vector field $X$ is defined by
\begin{equation}
I=\xi L+(\alpha-\dot{a}\xi)\frac{\partial L}{\partial \dot{a}}+(\beta-\dot{\phi}\xi)\frac{\partial L}{\partial \dot{\phi}}+(\gamma-\dot{\psi}\xi)\frac{\partial L}{\partial \dot{\psi}}-G.\label{13}
\end{equation}

Now if we assume that the Lagrangian (\ref{8}) admits the Noether symmetry on the tangent space $(a, \dot{a}, \phi, \dot{\phi}, \psi, \dot{\psi})$ then we get the following set of partial differential equations:
\begin{eqnarray}
\xi_{a}=\xi_{\phi}=\xi_{\psi}&=&0,\label{14}\\
-3\alpha a^2 V(\phi)-\beta a^3V'(\phi)-a^3V(\phi)\xi_t&=&G_t,\label{15}\\
6a\alpha_t=G_a,~-a^3\beta_t=G_{\phi},~-a^3F(\phi)\gamma_t&=&G_{\psi},\label{16}\\
-3\alpha-6a\frac{\partial \alpha}{\partial a}+3a \frac{\partial \xi}{\partial t}&=&0,\label{17}\\
\frac{3}{2}\alpha a^2+a^3 \frac{\partial \beta} {\partial \phi}-\frac{1}{2} a^3 \frac{\partial \xi}{\partial t}&=&0, \label{18}\\
\frac{3}{2}\alpha a^2F(\phi)+\frac{a^3}{2}\beta F'(\phi)+a^3F(\phi)\frac{\partial \gamma}{\partial \psi}-\frac{1}{2} a^3 F(\phi)\frac{\partial \xi}{\partial t}&=&0,\label{19}\\
-6a\frac{\partial \alpha}{\partial \phi}+a^3\frac{\partial \beta}{\partial a}&=&0,\label{20}\\
-6a\frac{\partial \alpha}{\partial \psi}+a^3F(\phi)\frac{\partial \gamma}{\partial a}&=&0,\label{21}\\
a^3\frac{\partial \beta}{\partial \psi}+a^3F(\phi)\frac{\partial \gamma}{\partial \phi}&=&0.\label{22}
\end{eqnarray}

Now the existence of the Noether symmetry demands the co-effitient of the infinitesimal generator (i.e., $\alpha, \beta, \gamma$) to satisfy the overdetermined set of partial differential equations. Thus, by using the method of separation of variable
i.e.,
\begin{eqnarray}
\alpha&=&\alpha_1(a) \alpha_2(\phi)\alpha_3(\psi),\nonumber\\
\beta&=&\beta_1(a)\beta_2(\phi)\beta_3(\psi),\nonumber\\
\gamma&=&\gamma_1(a)\gamma_2(\phi)\gamma_3(\psi),\nonumber
\end{eqnarray}
we get the explicit form of $\alpha, \beta$ and $\gamma$ which is given by
\begin{eqnarray}
\alpha&=&\frac{c_{\alpha}e^{m\phi}+c_{\beta}e^{-m\phi}}{a^{\frac{1}{2}}},\label{23}\\
\beta&=&\frac{-4m}{a^{\frac{3}{2}}}\bigg(c_{\alpha}e^{m\phi}-c_{\beta}e^{-m\phi}\bigg),\label{24}\\
\gamma&=&\gamma_0,~~~\xi=m_3t+m_4,\label{25}
\end{eqnarray}
where $c_{\alpha}, c_{\beta}, m_3, m_4$ and $\gamma_0$ are the arbitrary constants with $m^2=\frac{3}{8}$.

Also from Eqs. (\ref{15}) and (\ref{19}), we get the unknown potential function and the coupling function which depend on the scalar field ($\phi$) having the expression
\begin{eqnarray}
V(\phi)&=&V_0\bigg(c_{\alpha}e^{m\phi}-c_{\beta}e^{-m\phi}\bigg)^2,\label{26}\\
F(\phi)&=&F_0\bigg(c_{\alpha}e^{m\phi}-c_{\beta}e^{-m\phi}\bigg)^2,\label{27}
\end{eqnarray}
where $V_0, F_0$ are the integration constants. Hence on the tangent space, the infinitesimal generator turns out to be
\begin{eqnarray}
&X_1&=t\frac{\partial}{\partial t}+\psi \frac{\partial} {\partial \psi}-\frac{a}{3}\frac{\partial}{\partial a},\nonumber\\
&X_2&=\frac{\partial}{\partial t},\nonumber\\
&X_3&=\frac{\partial}{\partial \psi},\nonumber\\
&X_4&=\frac{2}{a^{\frac{1}{2}}}\cosh (m\phi)\frac{\partial}{\partial a}-\frac{8m}{a^{\frac{3}{2}}}\sinh (m\phi)\frac{\partial}{\partial \phi}.\label{28}
\end{eqnarray}
 From the Lagrangian in Eq. (\ref{8}), one may define the kinetic metric as
\begin{equation}
ds_{k}^2=-6a~ da^2+a^3~ d\phi^2+4F_0^2a^3\sinh^2 \phi ~d\psi^2\nonumber
\end{equation}
and the effective potential $V_{eff} (\phi)=4V_0a^3\sinh^2 \phi$. By choosing $u=a^{\frac{3}{2}}$, the above kinetic metric can be written as (with suitable scaling)
\begin{equation}
ds_{k}^2=-du^2+u^2(d\phi^2+\sinh^2 \phi ~d\psi^2)\nonumber
\end{equation}
Here the 2D metric $h_{AB}$ with $d\sigma^2=d \phi^2+\sinh^2 \phi ~d\psi^2$ corresponds to the representation of the $so(3)$ Lie Algebra. Further, it is possible to have a different representation of the $so(3)$ Lie algebra so that the above 2D metric can be written as \cite{palia}
\begin{equation}
h_{AB}=
\begin{pmatrix}
	1 & 0 \\
	0 & e^{2\phi}
\end{pmatrix}\nonumber
\end{equation}
Thus, the model can be reduced to describe hyperbolic inflation \cite{c}.

\section{Analytical Solution}
In this section, our aim is to find the exact cosmological solutions of this multi-scalar field model. As the existence of the Noether symmetry implies that there exists a cyclic co-ordiante, so we shall now turn our attention to find a new coordiante system in such a way such that one of the variables becomes cyclic. So by point transformation $(t, a, \phi, \psi)\rightarrow (s, u, v, w)$, the vector field $X$ transformed into
\begin{eqnarray}
\overrightarrow{X}_T&=&(i_{\overrightarrow{X}}dt)\frac{\partial}{\partial t}+(i_{\overrightarrow{X}}du)\frac{\partial}{\partial u}+(i_{\overrightarrow{X}}dv)\frac{\partial}{\partial v}+(i_{\overrightarrow{X}}dw)\frac{\partial}{\partial w}\nonumber\\&+&\Bigg(\frac{d}{dt}(i_{\overrightarrow{X}}du)\Bigg)\frac{d}{d\dot u}+\Bigg(\frac{d}{dt}(i_{\overrightarrow{X}}dv)\Bigg)\frac{d}{d\dot v}+\Bigg(\frac{d}{dt}(i_{\overrightarrow{X}}dw)\Bigg)\frac{d}{d\dot{w}},\label{29}
\end{eqnarray}
such that $i_{_{X_4}}ds=0, i_{_{X_4}}du=0, i_{_{X_4}}dv=0$ and $i_{_{X_4}}dw=1$, i.e.,
\begin{eqnarray}
\xi\frac{\partial s(t, a, \phi, \psi)}{\partial t}+\alpha\frac{\partial s(t, a, \phi, \psi)}{\partial a}+\beta\frac{\partial s(t, a, \phi, \psi)}{\partial \phi}+\gamma\frac{\partial s(t, a, \phi, \psi)}{\partial \psi}&=&0,\nonumber\\
\xi\frac{\partial u(t, a, \phi, \psi)}{\partial t}+\alpha\frac{\partial u(t, a, \phi, \psi)}{\partial a}+\beta\frac{\partial u(t, a, \phi, \psi)}{\partial \phi}+\gamma\frac{\partial u(t, a, \phi, \psi)}{\partial \psi}&=&0,\nonumber\\
\xi\frac{\partial v(t, a, \phi, \psi)}{\partial t}+\alpha\frac{\partial v(t, a, \phi, \psi)}{\partial a}+\beta\frac{\partial v(t, a, \phi, \psi)}{\partial \phi}+\gamma\frac{\partial v(t, a, \phi, \psi)}{\partial \psi}&=&0,\nonumber\\
\xi\frac{\partial w(t, a, \phi, \psi)}{\partial t}+\alpha\frac{\partial w(t, a, \phi, \psi)}{\partial a}+\beta\frac{\partial w(t, a, \phi, \psi)}{\partial \phi}+\gamma\frac{\partial w(t, a, \phi, \psi)}{\partial \psi}&=&1.\label{30}
\end{eqnarray}
Now solving Eq. (\ref{30}), one can get the form of scale factor ``$a$'' and the scale ``$\phi$'' and ``$\psi$'' in terms of new coordiantes (choosing $c_{\alpha}=c_{\beta}$)
\begin{eqnarray}
u&=&a^{\frac{3}{2}}\sinh(m\phi),\nonumber\\
v&=&\psi,\nonumber\\
w&=&\frac{2}{3}a^{\frac{3}{2}}\cosh(m \phi),\nonumber\\
s&=&t.\label{31}
\end{eqnarray}
As a consequence, the transformed Lagrangian takes the form
\begin{equation}
L=\frac{4}{3}\dot{u}^2-3\dot{w}^2+f_0u^2\dot{v}^2+4v_0u^2,\label{32}
\end{equation}
and the conserved energy in terms of new variable look like
\begin{equation}
E=\frac{4}{3}\dot{u}^2-3\dot{w}^2+f_0u^2\dot{v}^2-4v_0u^2.\label{33}
\end{equation}
Hence, the Euler-Lagrange equation in the new coordiantes system, using Eq. (\ref{32}), leads to
\begin{eqnarray}
4\ddot{u}-3f_0u\dot{v}^2-12uv_0&=&0\nonumber\\
2f_0u^2\dot{v}&=&\mbox{constant}\nonumber\\
\ddot{w}&=&0\label{34}
\end{eqnarray}
Thus by solving (\ref{34}), we get the explicit form of the scale factor of the Universe, the scalar fields, the potential and coupling function of the Universe as
\begin{figure}[h!]
	\centering
	\includegraphics[width=0.5\textwidth]{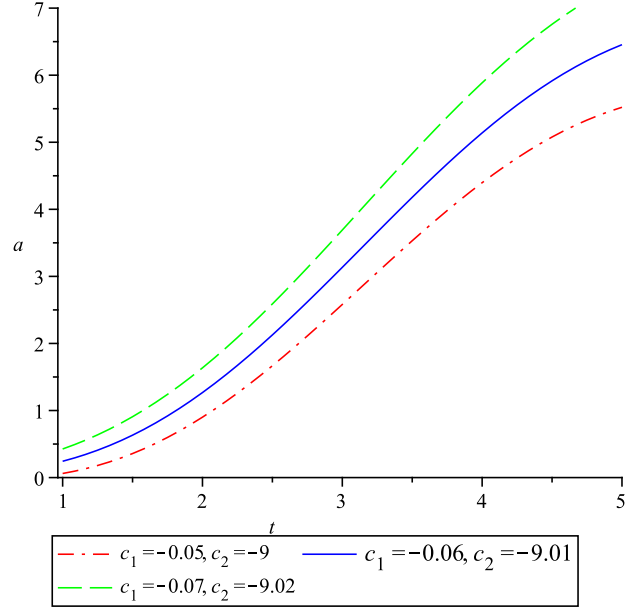}\\
	\caption{Graphical representation of the scale factor with respect to cosmic time $t$}
	\label{fig1}
\end{figure}

\begin{figure}[h!]
	\centering
	\includegraphics[width=0.5\textwidth]{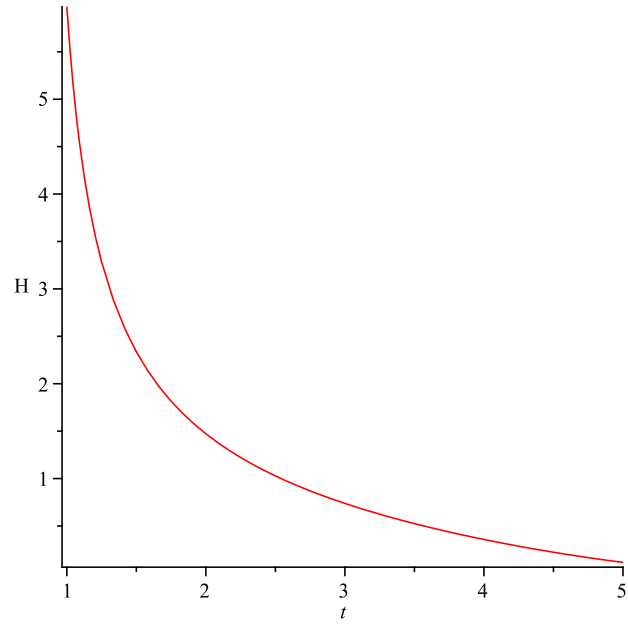}\\
	\caption{Represents the Hubble parameter with respect to cosmic time $t$}
	\label{fig2}
\end{figure}
\begin{figure}[h!]
	\centering
	\includegraphics[width=0.5\textwidth]{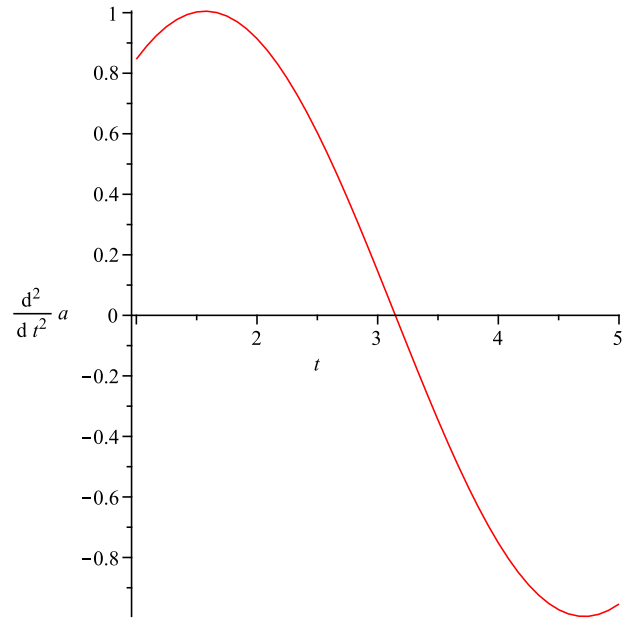}\\
	\caption{presents the acceleration parameter with respect to cosmic time $t$}
	\label{fig3}
\end{figure}
\begin{eqnarray}
a&=&\bigg[\bigg(c_1t+c_2\bigg)^2-\sin t-c_2^2\bigg]^{\frac{1}{3}},\nonumber\\
\phi&=&\tanh^{-1}\bigg\{\frac{\sqrt{\sin t+c_2^2}}{(c_1t+c_2)}\bigg\},\nonumber\\
\psi&=&\tan^{-1}\bigg\{\frac{s\tan (\frac{t}{2})+2}{\sqrt{s^2-1}} \bigg\},\nonumber\\
V(\phi)&=&V_0 \sinh^{2}\bigg[\tanh^{-1}\bigg\{\frac{\sqrt{\sin t+c_2^2}}{(c_1t+c_2)}\bigg\}\bigg],\nonumber\\
F(\phi)&=&F_0 \sinh^{2}\bigg[\tanh^{-1}\bigg\{\frac{\sqrt{\sin t+c_2^2}}{(c_1t+c_2)}\bigg\}\bigg],\label{35}
\end{eqnarray}
with $c_1, c_2, s, V_0, F_0$ as integration constants. These free parameters are related to the present value of the scale factor and Hubble parameter and also to the conserved Noetherian Charge. But due to complicated form of the solution, it is not possible to identify them in the above solution.

\section{Concluding Remarks}
This work is an example where a complicated coupled cosmological model has been solved analytically using symmetry analysis. The application of Noether symmetry to any physical system has two-fold advantages, namely (i) the symmetry conditions determine the symmetry vector field as well as any unknown parameter or function in the system (instead of choosing phenomenologically), (ii) using suitable transformation in the augmented space (choosing a cyclic coordinate) the Lagrangian as well as the evolution equations are simplified to a great extent so that analytic solutions can be obtained. A  similar two-scalar field cosmology has been studied by Paliathanasis and Tsamparlis \cite{palia}, where the scalar fields interact both in the kinetic part and the potential. But in this work, the two scalar fields are minimally coupled to gravity with a mixed kinetic term. In Ref. \cite{palia}, the authors have considered the Noether point symmetries corresponding to the elements of the homothetic group of the kinetic metric corresponding to the Lagrangian while in this work Noether symmetry vector is evaluated using Noether's theorem and then making a transformation in the augmented space (using the symmetry vector) the Lagrangian is simplified to a great extent by identifying a cyclic variable and the cosmological solution is obtained.

In the previous section, a coupled two-scalar field cosmological model has been solved completely using Noether symmetry analysis and the cosmological parameters, namely, the scale factor, Hubble parameter and the acceleration parameter have been shown graphically in Figs (\ref{fig1}--\ref{fig3}). The figures show that the present cosmological model is an expanding model of the Universe in an accelerating-decelerating eras of evolutions. Also, the Hubble parameter gradually decreases with time in accordance with observational evidences and it vanishes in finite time. Also, the model is useful for studying hyperbolic inflation in the early phase (after big-bang). Further, from the point of view of stability criterion, we have analyzed the solution for different choices of the parameter involved with differs infinitesimally (see Fig.(\ref{fig1})). It is found that the solution also differs infinitesimally. So the solution can be consider as stable. Finally, one may conclude that symmetry analysis is a powerful mathematical tool which is useful for analyzing complicated phenomenological cosmological model completely.
%$\begin{figure}
%	\centering
%	\includegraphics[width=0.5\textwidth]{f21.eps}\\
%	\caption{Hubble parameter}
%	\label{fig21}
%\end{figure}

%\begin{figure}
%	\centering
%	\includegraphics[width=0.5\textwidth]{f3.eps}%\\
	%\caption{Scale factor}
%	\label{fig3}
%\end{figure}
\section*{Acknowledgments}
 S.C. thanks Science and Engineering Research Board (SERB) for awarding
MATRICS Research Grant support (File No. MTR/2017/000407) and Inter
University Center for Astronomy and Astrophysics (IUCAA), Pune, India for their
warm hospitality as a part of the work was done during a visit.
 %%%%%%%%%%%%%%%%%%%%%%%%%%%%%%%%%%%%%%%%


\frenchspacing
 \begin{thebibliography}{58}
 
 \bibitem{r1} S. J. Perlmutter et al., {\it Astrophys. J.} {\bf 517}, 565 (1999).
 
\bibitem{o1} SDSS Collab. (D. J. Eisenstein et al.) , {\it Astrophys. J.} {\bf 633}, 560 (2005).

 \bibitem{o2} WMAP Collab. (C. L. Bennett et al.), {\it Astrophys. J. Suppl.} {\bf 208}, 20 (2013).
 %%%%%%%%%%%%%%%%%%%%%%%%%%%%%%%%
\bibitem{c1} A.D. Lindle, {\it Phys. Rev. D} {\bf 49}, 784 (1994).

 \bibitem{c2} E.J. Copeland, A.R. Liddle, D.H. Lyth, E.W. Steward and D. Wands, {\it Phys. Rev. D} {\bf 49}, 6410 (1994).
 
\bibitem{c3} S.A. Kim and A.R. Liddle, {\it Phys. Rev. D} {\bf 74}, 023513 (2006).

 \bibitem{c4} D. Wands, {\it Lect. Notes Phys.} {\bf 738}, 275 (2008).
 
\bibitem{c5} P. Carrilho, D. Mulryne, J. Ronaye and T. Tenkanen, {\it J. Cosmol. Astropart. Phys.} {\bf 06}, 032 (2018).

\bibitem{c6} P. Christodoulidis, D. Roest, E.I. Sfakianakis, \textit{J. Cosmol. Astropart. Phys.} \textbf{8}, 006 (2020).
 
 
 \bibitem{c7} K. Maeda, S. Mizuno and R. Tozuka, {\it Phys. Rev. D} {\bf 98}, 123530 (2018).
 
 \bibitem{c8} T. Kobayashi, O. Seto and T.H. Tatsuishi, {\it PTEP} {\bf 12}, 123B04 (2017).
 
\bibitem{c9} S.V. Chervon, {\it Quantum Matter} {\bf 2}, 71 (2013).

\bibitem{c10} Y.F. Cai, E.N. Saridakis, M.R. Setare and J.-Q. Xia, {\it Phys. Rep.} {\bf 493}, 1 (2010).

\bibitem{c11} E.N. Saridakis and J.W. Weller, {\it Phys. Rev. D} {\bf 81}, 123523 (2010).

\bibitem{c12} S.V. Chervon, {\it Russ. Phys. J.} {\bf 38}, 539 (1995).

\bibitem{c13} S.V Chervon, S.D. Maharaj, A. Beesman and A.S. Kubasov, {\it Gravit. Cosmol.} {\bf 20}, 176 (2014).

\bibitem{c14} S.V. Chervon, {\it Russ. Phys. J.} {\bf 39}, 139 (1996).

\bibitem{c15} S. V. Ketov, {\it Quantum Non-linear Sigma Models}  (Springer-Verlag, 2000).

\bibitem{c16} J. Lee, T.H. Lee, T. Moon and P. Oh, {\it Phys. Rev. D} {\bf 80}, 0656016 (2009)
%%%%%%%%%%%%%%%%%%%%%%%%%%%%%%%%%%%%%%
\bibitem{m1} C.-B. Chen, J. Soda {\it J. Cosmol. Astropart. Phys.} {\bf 09}, 026 (2021).
\bibitem{m2} A. Paliathanasis {\it Universe} {\bf 7}, 323 (2021) arXiv:2108.12154 [gr-qc].
\bibitem{m3} A. Giacomini, P.G.L. Leach, G. Leon, A. Paliathanasis, \textit{Eur. Phys. J. Plus} \textbf{136}, 1018 (2021).
\bibitem{m4} A. Paliathanasis, G. Leon arXiv:2105.03261 [gr-qc].
\bibitem{m5} A. Paliathanasis, G. Leon {\it Class. Quantum Grav}. {\bf 38}, 075013 (2021).
\bibitem{m6} A. Paliathanasis, G. Leon {\it Eur. Phys. J. C}. {\bf 80}, 840 (2020).
\bibitem{m7} A. Giacomini, E. González, G. Leon, A. Paliathanasis arXiv:2104.13649 [gr-qc].
\bibitem{m8} V. R. Ivanov, S. Yu. Vernov, \textit{Eur. Phys. J. C} \textbf{81}, 985 (2021).
\bibitem{m9} A. A. Coley, R.J. v. den Hoogen {\it 	Phys.Rev. D}. {\bf 60}, 023517 (2000).
\bibitem{m10} A. Giacomini, G. Leon, A. Paliathanasis, S. Pan {\it Eur. Phys. J. C}. {\bf 80}, 184 (2020).
\bibitem{m11} P. Christodoulidis {\it Eur. Phys. J. C}. {\bf 81}, 471 (2021).
%%%%%%%%%%%%%%%%%%%%%%%%%%%%%%%%%%%%%%%%%5
\bibitem{palia} A. Paliathanasis, M. Tsamparlis, {\it Phys. Rev. D} {\bf 90}, 043529
(2014).
 \bibitem{a} S.V. Chervon, I.V. Fomin, E.O. Pozdeeva, M. Sami, S.Yu. Vernov, {\it Phys. Rev. D} {\bf 100}, 063522
(2019).
\bibitem{1} V. Faraoni, S. Jose, S. Dussault 	arXiv:2107.12488 [gr-qc] 
%%%%%%%%%%%%%%%%%%%%%%%%%%%%%%%%%%%%555
 \bibitem{r2} N. Dimakis, A. Paliathanasis, Petros A. Terzis, T. Christodoulakis, {\it Eur. Phys. J. C} {\bf 79}, 618 (2019).
 %%%%%%%%%%%%%%%%%%%%%%%%%%%%%%%%%%%%%
 \bibitem{b} A. Borowiec, A. Kozak (arXiv:2108.13324 [gr-qc]).
 %%%%%%%%%%%%%%%%%%%%%%%%%%%%%%%%
 \bibitem{n1} A. Paliathanasis, L. Karpathopoulos, A. Wojnar, S. Capozziello, {\it Eur. Phys. J. C} {\bf 76}, 225 (2016).
 
 \bibitem{n2} S. Capozziello, R. de Ritis, P. Scudellaro, {\it Int. J. Mod. Phys. D} {\bf 2}, 463 (1993).
 
 \bibitem{n3} S. Capozziello, G. Marmo, C. Rubano, P. Scudellaro, {\it Int. J. Mod. Phys. D} {\bf 6}, 491 (1997).
 
 \bibitem{n4} S. Capozziello, R. de Ritis, C. Rubano, P. Scudellaro, {\it Riv. Nuovo.Cim} {\bf 19}, 114 (1996).
 
 \bibitem{n5} S Basilakos, M Tsamparlis and A Paliathanasis, {\it Phys.Rev. D} {\bf 83}, 103512 (2011);
 A Paliathanasis, M Tsamparlis, S Basilakos and J. D. Barrow, {\it Phys.Rev. D} {\bf 91}, 123535 (2015).
 
\bibitem{n6}  S. Capozziello, F. Darabi and D. Vernieri, {\it Mod. Phys. Lett. A} {\bf 26}, 65 (2011).

\bibitem{n7} S. Capozziello and A. De. Felice, {\it J. Cosmol. Astropart. Phys.} {\bf 0808}, 016 (2008).
 %%%%%%%%%%%%%%%%%%%%%%%%%%%%%%5
 
 \bibitem{r3} K. F. Dialektopoulos, S.Capozziello, {\it Int. Jour. Geom. Meth. Mod. Phys.} {\bf 15}, 1840007 (2018)
  (arXiv:1808.03484 [gr-qc]).
  15
  \bibitem{r4} S. Basilakos, S. Capozziello, M. De Laurentis, A. Paliathanasis, M. Tsamparlis, {\it Phys. Rev. D} {\bf 88}, 103526
  (2013).
 
 %%%%%%%%%%%%%%%%%%%%
 %%%%%%%%%%%%%%%%%%%%%%%%%%%%
 


\bibitem{c} A. R. Brown, \textit{Phys. Rev. Lett.} \textbf{121}, 251601 (2018).


 \end{thebibliography}
\end{document}